\documentclass{aa}
\usepackage{natbib}
\bibpunct{(}{)}{;}{a}{}{,}  
\usepackage{graphicx}
\usepackage{txfonts}
\begin{document}

\title{Optical Spectroscopy of Active Galactic Nuclei in SA57\thanks{Based on observations made with the William Herschel Telescope (WHT), operated by the ING, and with the Italian Telescopio Nazionale Galileo (TNG), operated on the island of La Palma by the Fundaci—n Galileo Galilei of the INAF (Istituto Nazionale di Astrofisica), both at the Spanish Observatorio del Roque de los Muchachos of the Instituto de Astrofisica de Canarias.}}

\author{D. Trevese \inst{1}, V. Zitelli \inst{2}, F. Vagnetti \inst{3}, K. Boutsia \inst{1,4}, G.M. Stirpe\inst{2}}

   \institute{Dipartimento di Fisica, Universit\'a di Roma ``La Sapienza'', P.le A. Moro 2, I-00185 Roma (Italy) \\
   \email{dario.trevese@roma1.infn.it}
         \and
	  INAF - Osservatorio Astronomico di Bologna, via Ranzani 1, I-40127 Bologna (Italy)
         \and
	 Dipartimento di Fisica, Universit\'a di Roma ``Tor Vergata'', Via delle Ricerca Scientifica 1, I-00133 Roma (Italy)
         \and
	 European Southern Observatory, Karl-Schwarzschild-Stra\ss e 2, D-85748 Garching (Germany)
             }

 \date{}

\abstract
% context heading (optional) leave it empty if necessary 
{The cosmological evolution of X-ray-selected and optically selected Active Galactic Nuclei (AGNs)
show different behaviours  interpreted in terms of two different populations. The difference is evident 
mainly for low luminosity AGNs (LLAGNs), many of which are lost by optical photometric surveys.}
 % aims heading (mandatory) 
 {
 We are conducting a spectroscopical study of a composite sample of AGN candidates selected in SA57 following different searching techniques, to identify low luminosity AGNs and break down the sample
 into different classes of objects.
 }  
  % methods heading (mandatory)
{
AGN candidates were obtained through optical variability and/or X-ray emission.
Of special interest are the extended variable objects, which are
expected to be galaxies hosting LLAGNs.
 }
 % results heading (mandatory)
{Among the 26 classified objects a fair number (9)  show typical AGN spectra. 10 objects show Narrow Emission Line Galaxy spectra, and  in most of them (8/10) optical variability suggests the presence of LLAGNs.
}
% conclusions heading (optional), leave it empty if necessary 
{}


\authorrunning{D. Trevese et al.}
\titlerunning{Optical Spectroscopy of AGNs in SA57}
\maketitle
\section{Introduction}
In recent years a growing amount of evidence suggested the existence of a link between the evolution in cosmic time of galaxy and quasar (QSO) 
populations.  Theoretical work discusses the effect of galaxy merging on the nuclear activity, through an increment of the rate of accretion onto the massive black hole, hosted in (possibly all) galaxy nuclei \citep{kor95}. 
Observationally, the cosmic history of active galactic nuclei (AGNs) is deduced from the analysis of  optical and X-ray luminosity functions (LFs) and their redshift dependence. 
While optical observations imply that the maximum of the QSO/AGN number density occurs at $z_M \ga2$ independently of their absolute luminosity \citep{wolf03}, 
X-ray surveys indicate a "cosmic downsizing"  with the epoch of maximum density going from  $z_M\sim 1.5$, for bright objects
($L_X (2-10\,{\rm keV}) \sim 10^{45}  {\rm\, erg\, s}^{-1}$), to $z_M \sim 0.5$ for faint ones ($L_X (2-10\,{\rm keV})\sim 10^{42}  {\rm\, erg\, s}^{-1}$)   \citep{ued03,lafr05}. 
It has been suggested that this behaviour is a consequence of the existence of two distinct AGN populations: the first consisting of QSOs and brighter AGNs,
born at high $z$ from frequent galaxy merging in high density regions, and related to the red part of the bimodal galaxy distribution, and the second population made of smaller and gas-rich galaxies still providing material for feeding smaller black holes and energizing low luminosity AGNs at later times through galaxy interactions \citep{cav07}. 
To evaluate sample completeness and selection effects,
 more detailed analysis of the LF evolution is necessary, in particular for low luminosity, optically selected AGNs, to accurately quantify the intrinsic evolution.
Brighter AGNs are detected in the optical band, mainly by their non-stellar colour, i.e. their position outside the "stellar locus" in  colour space. 
Fainter AGNs cannot be detected in the same way, since the observed spectral energy distribution (SED) is dominated by the host galaxy and in general is non-stellar independently of the presence of an active nucleus.
Thus,  probing the cosmic downsizing  in optical  samples requires a different selection technique. 
Recently \citet{bon07} have selected  a complete, volume limited sample of 130 type 1 AGNs from the catalogue of 150000 spectra obtained by the VIMOS-VLT Deep Survey,  \citep{lefe05}. 
From this sample they have found a first evidence
that the peak in density of lower luminosity  type 1 AGNs is  progressively shifted towards lower redshifts.

Another way to select  AGNs  is based on the detection of their variability. The method, which was proposed for the first time by \citet{vdb73},
has different completeness and reliability depending on 
the accuracy of photometric measurements and the distribution of sampling times.
It also depends on the total duration  of the observing campaign, since the r.m.s. variation increases 
with the lag between observations \citep{bono79,gial91,t94,vand04,devr05}.
It has been applied in the past to various data-sets \citep[e.g.,][]{t89,cris90,ver95}. It is particularly interesting in the case of low luminosity AGNs (LLAGNs),
where  the image is not point-like and the colour selection fails. This search for  ``variable galaxies" has been experimented by \citet[][hereinafter BTK]{btk98} in the field of Selected Area 57 (SA57) where other techniques like the selection by colour and by the absence of proper motion were also applied \citep{kc81,kkc86}, and a few brighter candidates were confirmed spectroscopically. 
A similar procedure was applied to multi-epoch Hubble Space Telescope images  by \citet{sara03,sara06}, who have also shown that some of the
variability selected AGNs are not detected in X-rays.
We have observed SA57 in X-rays with XMM-Newton (for 67 ks) and obtained a catalogue of 140
AGN candidates \citep{trev07}, 98 of which are identified with optical images mainly from the KPNO survey of  that field \citep{kron80,koo86}.
Candidates for spectroscopy are either selected from optical variability or from  X-ray emission.
In this paper we describe the result of a first part of this spectroscopic campaign.

The paper is organized as follows. \S 2 describes spectroscopic observations and data reduction, \S 3 describes the results and \S 4 contains a discussion of the results.

We adopt the concordance cosmology,
$H_o=75$ km s$^{-1}$Mpc$^{-1}$, $\Omega_m=0.3$,  $\Omega_{\Lambda}=0.7$,  throughout the paper.

\section{Observations and data reduction}
Observations were carried out with the fiber-fed multi-object spectrograph AF2/WYFFOS at the 4.2m William Herschel Telescope (WHT),
La Palma (Canary Islands - Spain) , on 28-29 April 2006. In a single, partially clear night on April 1 2006 we obtained also the spectrum of
one object with DOLORES at Telescopio Nazionale Galileo (TNG) , La Palma. The 1.6 arcsec diameter of the AF2/WYFFOS fibers requires a relatively accurate positioning.
For this reason we re-computed the absolute $\alpha,\delta$ in the following way.
We started from the catalogue of the Kitt Peak National Observatory (KPNO) survey of SA57 \citep{kron80,koo86},
based on $U, B_J, F,N$ photometry taken almost yearly between 1974 and 1989. 
We cross-correlated the object positions with position in the USNO-A2.0 catalogue.
After a 2-$\sigma$ rejection, the IRAF {\it ccmap} utility
provides a 4th order coordinate transformation based on 446 objects spread over the field,
with $<$0.2 arcsec r.m.s. deviation in both $\alpha$ and $\delta$, with respect to USNO-A2.0.
The  2-$\sigma$ rejection eliminates most of the high proper motion objects whose positions underwent significant 
changes between the epochs of the KPNO master catalogue (1974) and the USNO-A2.0 catalogue (1956).
Proper motions in the field were published by \citet{maj92}. This allowed us to select the guiding stars
among low proper motion ($< 1.2$ arcsec/yr) objects and compute their position at the epoch of observation. The result was an accurate centering of all  of the selected guiding stars.
Candidates were selected from three different lists:  i) variable extended candidates from BTK 
(details on objects fainter than 22.5 were not reported in the BTK paper);  ii) point-like variable objects from \citet{t89},
not yet observed spectroscopically ; iii) objects from a subsample of 98 among 140 X-ray detected 
objects from \citet{trev07} (see notes to Tab. \ref{Tab1} in the next section). 
Some previously observed objects were automatically included by the fiber positioning software in those cases 
in which some of the candidates were not observable due to the crowding of the field.  
Observations were carried out for two successive nights in April 2006. We adopted two different configurations,
such that fainter objects had a total exposure of 12 hours,  i.e. were present in both configurations and were observed for two nights,  while  brighter objects were observed for 6 hours during the first or the second night.
Between 15 and 20  fibers spread over the field were dedicated to the sky spectrum.
We used the grating R158B providing a dispersion of 2.0 \AA/px, the detector was the 2-chip EEV mosaic,  read with  $2 \times 2$ binning
and the resolution was 16 \AA.
Individual observing blocks lasted 30 minutes each and were co-added to form images of 2 hours of total exposure.
For each image we recorded also arc spectra for wavelength calibration.
Data reduction was carried out with standard IRAF procedures. Spectra were extracted from each (2 hours) exposure and were calibrated in wavelength. 
%No automatic pixel rejection was adopted to remove ``hot pixels''. 
%Instead, whenever an extremely narrow line appeared in a spectrum, a check on the other exposures and on the image was performed to identify the ``hot pixels''.
% spectra were co-added to form the final spectra @.
An average, and high signal to noise,
spectrum of the sky was obtained by combining several sky spectra. Sky subtraction was performed after a normalization of the average sky spectrum to minimize the residuals in correspondence of the main sky emission lines. In some cases, a single sky spectrum from a fibre close to the object was subtracted instead. The spectrum of the object NSER 16338 (see \S 3) was also obtained with DOLORES at TNG, in long slit  mode. It was obtained in a single 40 min exposure, instead of the 160 min planned, due to bad weather conditions.

\section{Results}
In the following we show the optical spectra (Figs. \ref{Fig1a}-\ref{Fig1e} and \ref{Fig2}) and provide notes on individual objects.

\renewcommand{\thefigure}{1\alph{figure}}
\begin{figure}
\centering
\resizebox{\hsize}{!}{\includegraphics{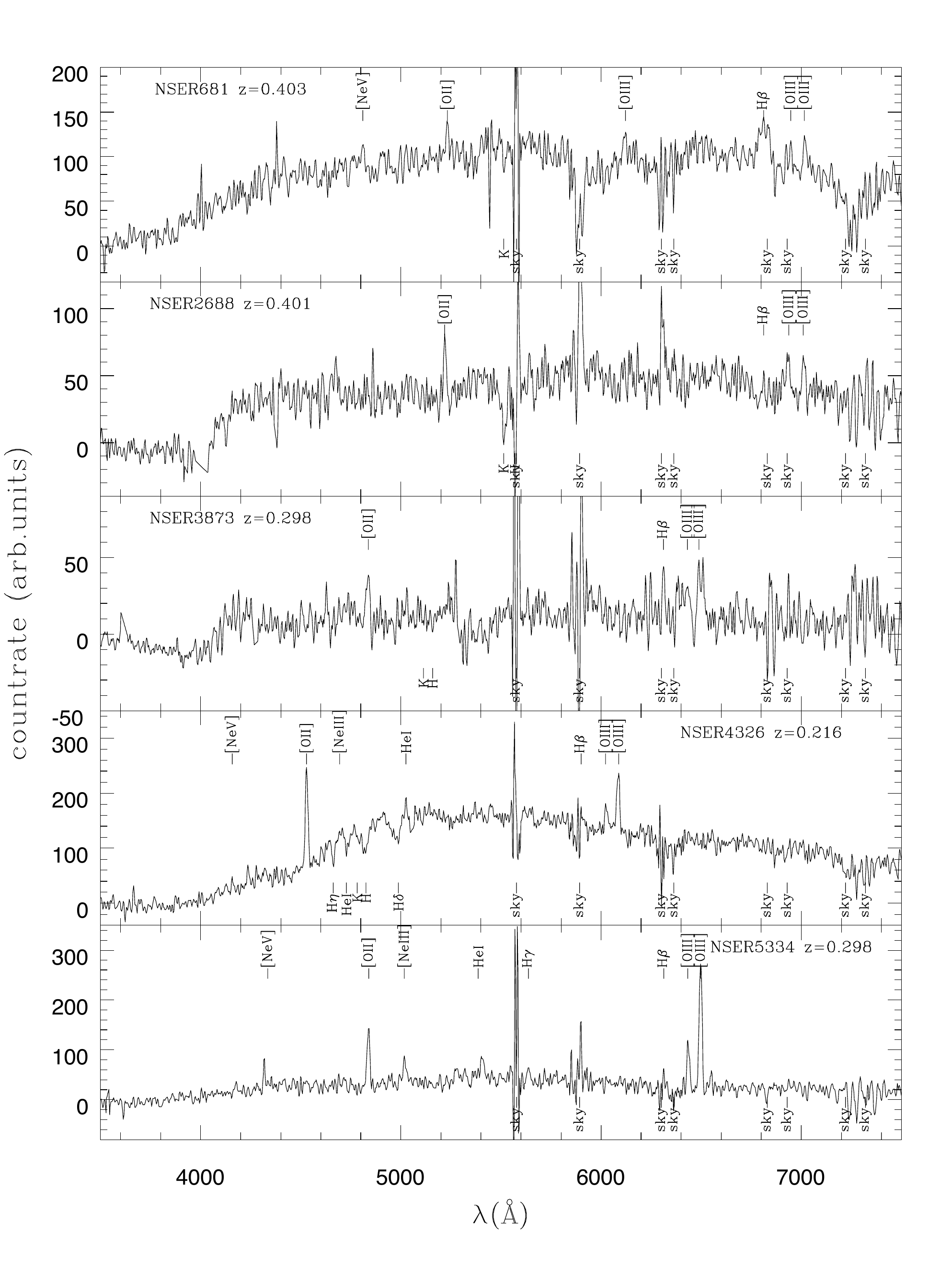}}
%\caption{(continued on the next pages@). Spectra of objects observed with WYFFOS at WHT.}
\caption{Spectra of objects observed with WYFFOS at WHT.}
         \label{Fig1a}
\end{figure}

\begin{figure}
\centering
\resizebox{\hsize}{!}{\includegraphics{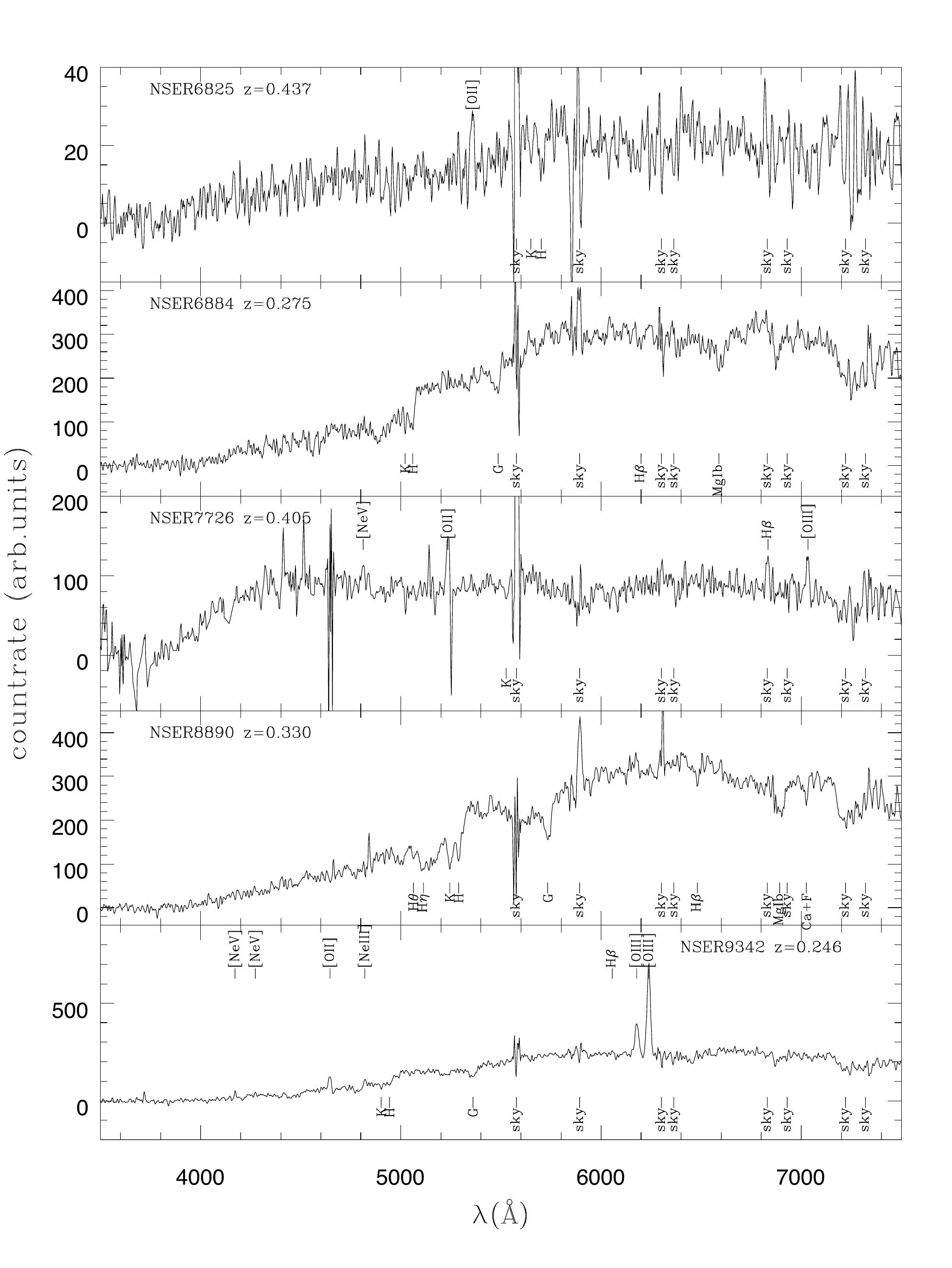}}
\caption{(continued).}
         \label{Fig1b}
\end{figure}

\begin{figure}
\centering
\resizebox{\hsize}{!}{\includegraphics{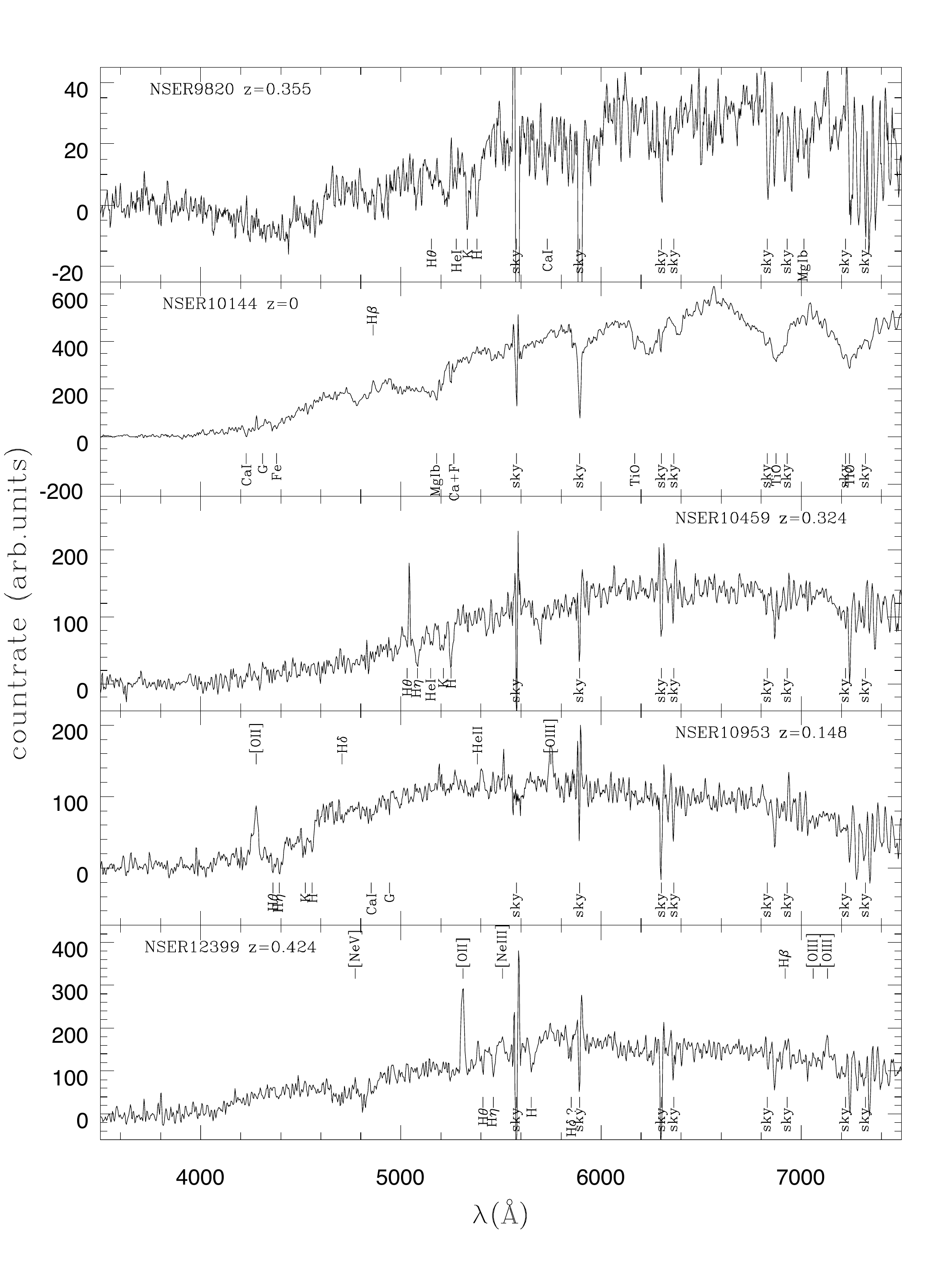}}
\caption{(continued).}
         \label{Fig1c}
\end{figure}

\begin{figure}
\centering
\resizebox{\hsize}{!}{\includegraphics{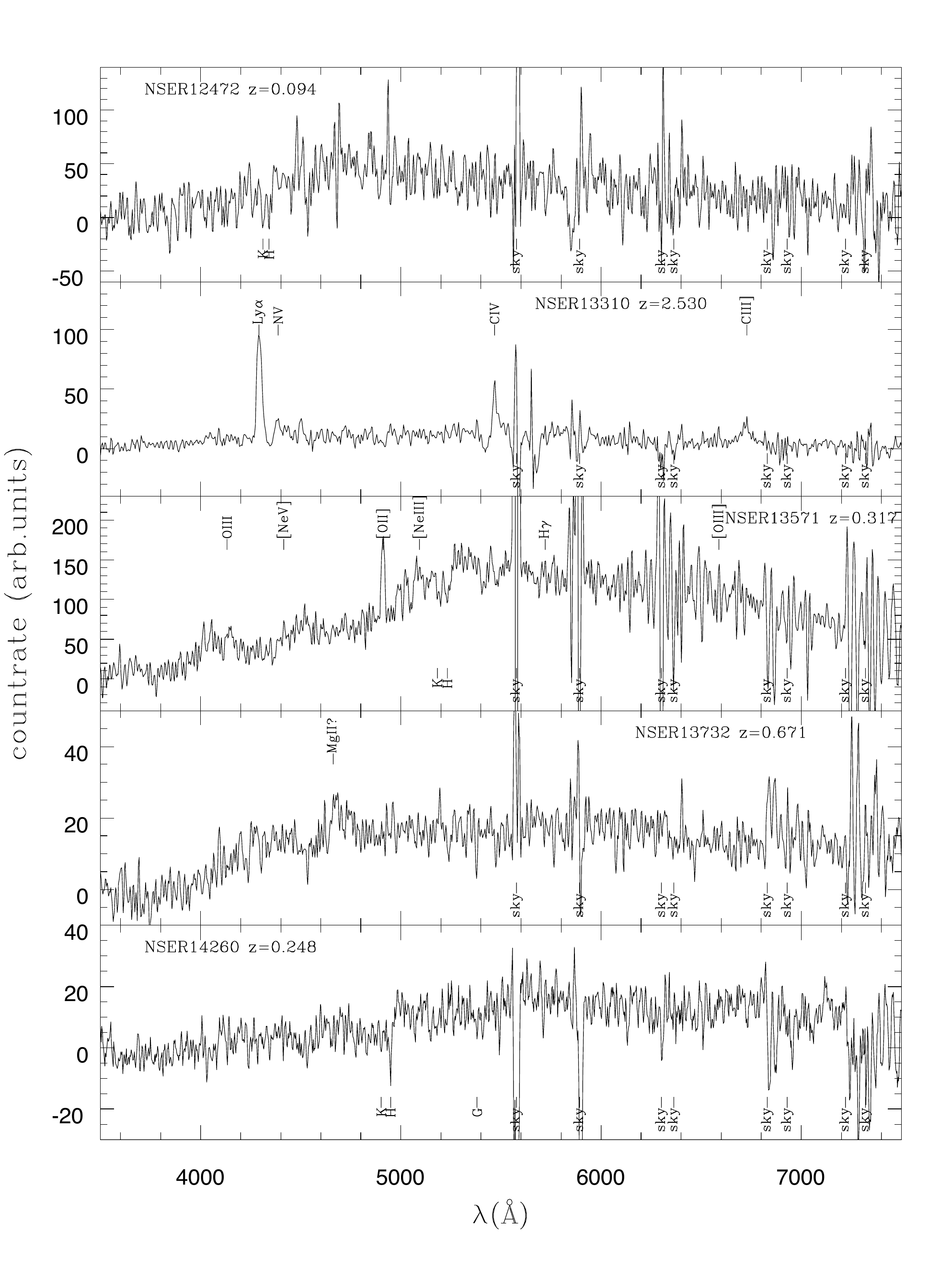}}
\caption{(continued).}
         \label{Fig1d}
\end{figure}

\begin{figure}
\centering
\resizebox{\hsize}{!}{\includegraphics{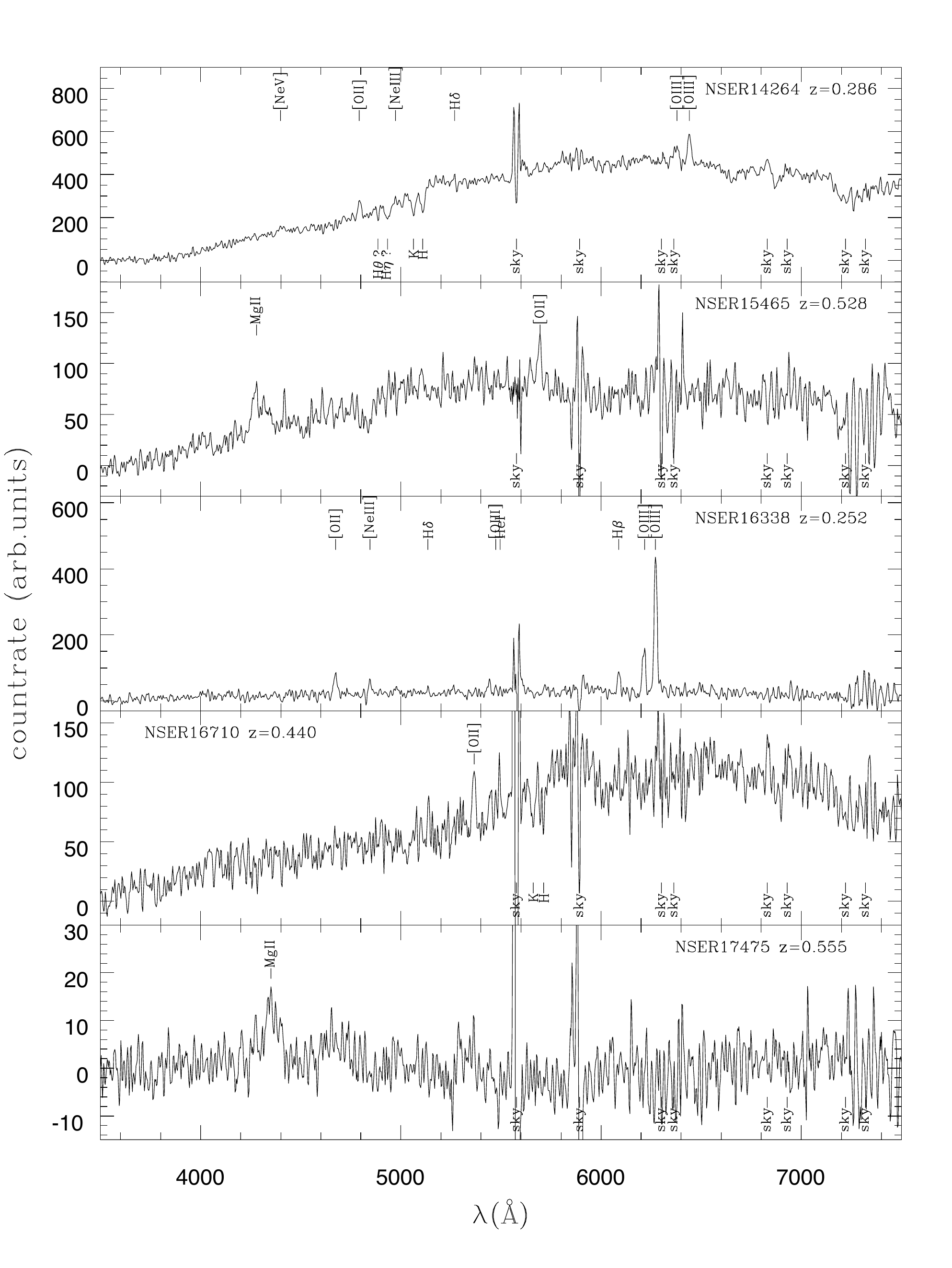}}
\caption{(continued).}
         \label{Fig1e}
\end{figure}

%%%%%%%%% Fig 2
\renewcommand{\thefigure}{\arabic{figure}}
\setcounter{figure}{1}
\begin{figure}
\centering
\resizebox{\hsize}{!}{\includegraphics{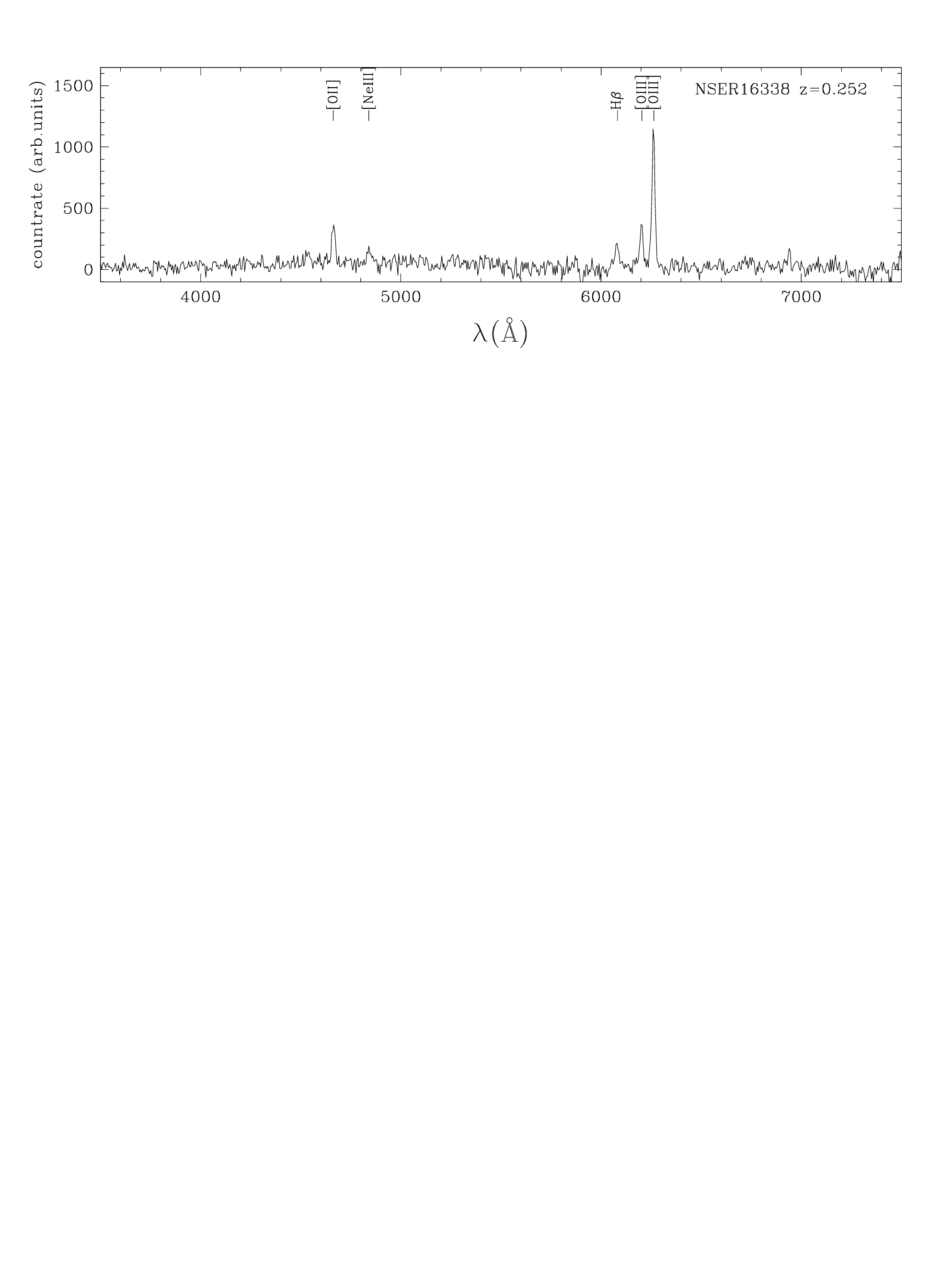}}
\vspace{-9cm}
\caption{Spectrum of NSER 16338 obtained with DOLORES at TNG}
         \label{Fig2}
\end{figure}

\subsection{Optical  spectra}

In Table \ref{Tab1}  we list the objects.
The meaning of the columns is as follows:
{\it Column 1}:  serial number NSER\footnote{Notice that in some publications \citep[BTK]{munn97} the same objects are designed  by a serial number equal to (NSER+100,000)} in the KPNO survey \citep{kron80,koo86};
{\it Column 2 and 3}: J2000 coordinates;
{\it Column 4 and 5}:  $B_J$ and F magnitudes;
{\it Column 4}: note on variability selection;
{\it Column 6}: identification number in the X-ray catalogue of \citet{trev07};
{\it Column 7}: notes on the identification;
{\it Column 8}: redshift;
{\it Column 9}: notes on redshift determination;
{\it Column 10}: classification.
 
\begin{table*}
\begin{center}
\caption{ \label{Tab1} Observed objects}
\begin{tabular}{rcccccccccl}
\hline
 NSER      & RA(2000)   & DEC(2000)    & $B_J$  & $F$     &nv$^a$&SA57X$^b$ &ni$^c$&$z$&nz$^d$& classification$^e$ \\
\hline
\hline
  681      & 13 08 05.91& +29 06 53.5  & 19.955 &   19.261  & EV & -    &   & 0.403  &c  & BLAGN \\
 2688      & 13 09 42.82& +29 11 24.7  & 22.280 &   20.868  & EV & -    &   & 0.401  &   & NELG \\
 3873      & 13 09 36.84& +29 13 23.9  & 22.345 &   21.503  & EV & -    &   & 0.298  &   & NELG \\
 4326      & 13 08 53.67& +29 14 06.1  & 20.867 &   20.172  & EV & -    &   & 0.216  &   & NELG \\
 5334      & 13 09 44.08& +29 15 47.0  & 22.235 &   21.249  & EV & -    &   & 0.298  &e  & NLAGN \\
 6825      & 13 08 22.19& +29 18 07.8  & 22.920 &   21.262  &    & 37   &   & 0.437  &   & NELG \\
 6884      & 13 08 13.52& +29 18 12.5  & 20.833 &   19.006  &    & 38   &   & 0.275  &a  & XBONG \\
 7726      & 13 08 34.17& +29 19 27.0  & 21.446 &   20.409  & EV & -    &   & 0.405  &   & NELG \\
 8553      & 13 07 13.93& +29 20 42.1  & 20.946 &   19.675  & EV & -    &   & 0.297  &a,c& NELG \\
 8890      & 13 08 55.51& +29 21 10.5  & 20.964 &   19.051  &    & 61   &   & 0.330  &a& galaxy \\
 9342      & 13 08 56.78& +29 21 50.5  & 20.885 &   19.140  &    & 66   & A & 0.246  &   & NLAGN \\
 9820      & 13 08 56.74& +29 22 29.0  & 21.868 &   20.097  &    & 71   & M & 0.355  &a  & XBONG \\
10144      & 13 09 00.18& +29 22 58.9  & 20.180  &   18.471  &    & 81   &   &  -     &   & star \\
10459      & 13 09 34.63& +29 23 28.1  & 21.366 &   19.682  & EV & -    &   & 0.324  &a,c& galaxy \\
10953      & 13 08 50.34& +29 24 15.0  & 20.960 &   20.071  &    & 90   & A & 0.148  &   & NELG \\
12399      & 13 09 51.62& +29 26 17.9  & 21.930 &   20.710  & PV & -    &   & 0.424  &   & NELG \\
12472      & 13 07 29.20& +29 26 25.4  & 22.578 &   20.933  &    & 102  &   & 0.094  &e,u& galaxy \\  
13310      & 13 07 56.71& +29 27 38.0  & 21.656 &   21.074  &    & 109  & M & 2.530  &e  & BLAGN \\
13571      & 13 09 49.30& +29 28 00.7  & 21.713 &   20.677  & EV & -    &   & 0.317  &   & NELG \\
13732      & 13 08 24.04& +29 28 19.2  & 22.763 &   22.640  &    & 115  &   & 0.671  &e,u& BLAGN \\
14260      & 13 07 58.37& +29 29 08.2  & 22.223 &   20.718  & EV & -    &   & 0.248  &a,u& galaxy\\
14264      & 13 08 03.40& +29 29 08.8  & 20.391 &   18.782  & EV & 120  &   & 0.286  & & NLAGN \\
15465      & 13 09 17.09& +29 31 04.3  & 21.886 &   21.151  & PV & 127  &   & 0.528  &e  & BLAGN \\
16338      & 13 07 30.34& +29 32 22.5  & 22.719 &   21.957  & EV & -    &   & 0.252  &e  & NLAGN \\
16710      & 13 09 03.87& +29 33 06.3  & 21.690 &   20.119  & EV & -    &   & 0.440  &u  & NELG \\
17475      & 13 07 53.14& +29 34 17.0  & 22.551 &   22.410  & PV & -    &   & 0.555  &e,u& BLAGN \\
\hline
\hline
\end{tabular}
\end{center}
$^a$ notes on variability selection. EV: extended variable; PV: pointlike variable \\
$^b$ X-ray catalog number following \citet{trev07} \\
$^c$ notes on optical identification. M: marginal; A: ambiguous \\
$^d$ notes on redshift. e: only emission; a: only absorption; c: confirmed redshift (see BTK); u: uncertain\\ 
$^e$ BLAGN: Broad Line AGN; NLAGN: Narrow Line AGN; NELG: Narrow Emission Line Galaxy;
XBONG: X-ray Bright Optically Normal Galaxy; galaxy: galactic spectrum with only absorption lines.
\end{table*}
 
%The classification is @@
%Cndidates selected on the basis of variability are indicated ..@
%Candidates selected from the X-ray catalogue \citep{trev07}
The tentative classification reported in Table \ref{Tab1} is based on the appearence of the optical spectra.
A distinction among starburst galaxies, Seyfert 2's, and Low Ionization Narrow Emission Regions (LINERs) would require the knowledge of the ([NII] $\lambda$6583)/H${\alpha}$$\lambda$6563)
ratio which is out of our spectral range 
for $z\ga 0.4$. Still we  assume as Seyfert 2s the objects with  large
([OIII] $\lambda$5007/H$\beta$ $\lambda$4861)$\ga$ 3 \citep[see Fig. 1][]{vo87}.
These objects are classified as Narrow Line AGNs (NLAGNs). Where no broad lines are seen and
([OIII] $\lambda$5007/H$\beta$ $\lambda$4861) is not large enough, the object may be either a starburst galaxy or a LINER and we simply designate it as Narrow Emission Line Galaxy (NELG). 
Objects are classified as Broad Line AGNs whenever the broad components are detected. It should be noted, however, that
in the case of LLAGNs, this classification becomes rather "fuzzy", since the detectability of the broad lines depends on both the S/N ratio and the relative importance of the nuclear component respect to the host galaxy. Notice that this is, obviously, true for any LLAGN sample,  no matter how detected.
 
\subsection{Notes on individual objects}

\begin{itemize}
\item {\it NSER 681}: Broad-line AGN. The emission-line redshift $z_e=0.403$ is confirmed as previously found by BTK. We don't see the MgII$\lambda$2798 emission line. Instead, we reveal a broad H$\beta$. Ca II K $\lambda$3934 allows determination of an absorption-line redshift $z_a=0.402$.
\item {\it NSER 2688}: NELG. $z=0.401$ based on [OII]$\lambda$3727, H$\beta$, [OIII]$\lambda$4959 and $\lambda$5007, and on CaII K absorption. 
\item {\it NSER 3873}: NELG. $z=0.298$ based on [OII]$\lambda$3727, H$\beta$, [OIII]$\lambda$4959 and $\lambda$5007, and on CaII H and K absorption.
\item {\it NSER 4326}: NELG with a strong [OII]$\lambda$3727 and a relatively weak [OIII]$\lambda$5007. Emission and absorption features agree on a redshift $z=0.216$.
\item {\it NSER 5334}: Emission-line redshift $z_e=0.298$. The strong narrow emission lines [OII]$\lambda$3727, [OIII]$\lambda$4959 and $\lambda$5007, and the low H$\beta$/[OIII]$\lambda$5007 ratio favor a Seyfert 2 classification.
\item {\it NSER 6825}: $z=0.437$ based on [OII] $\lambda$3727 emission and on CaII K and H absorption. NELG.
\item {\it NSER 6884}: No emission features. $z_a=0.275$ based on CaII K and H and on H$\beta$ and MgIb $\lambda$5175.4. An Absorption-line Galaxy which can be classified XBONG due to its relatively high X-ray luminosity ($L_{2-10{\rm keV}}\la 10^{42}$ erg/s, cf. \citet{trev07}).
\item {\it NSER 7726}: Emission features and CaII K$\lambda$3934 absorption agree on a redshift determination $z=0.405$. Relatively strong [OII]$\lambda$3727 and a high  H$\beta/$[OIII]$\lambda$5007 ratio suggest this object is a NELG.
\item {\it NSER 8553}: We confirm the redshift $z=0.297$ by BTK. Our spectrum has worse S/N ratio and is not reported in Fig. 1.
\item {\it NSER 8890}: An Absorption-line galaxy with $z_a=0.330$ based on CaII K and H and on H$\eta$ and H$\theta$.
\item {\it NSER 9342}: A Seyfert 2 with redshift $z=0.264$ based on [OII]$\lambda$3727, strong [OIII]$\lambda$4959, $\lambda$5007, and on CaII H and K absorption. 
\item {\it NSER 9820}: No emission features. Absorption-line Galaxy with $z_a=0.355$ based on H$\theta$, HeI $\lambda$3889, CaII K and H, CaI $\lambda$4226.7 and MgIb $\lambda$5175.4. 
Its relatively high X-ray luminosity ($L_{2-10{\rm keV}}\la 10^{42}$ erg/s, cf. \citet{trev07}) puts it in the range of XBONGs.
\item {\it NSER 10144}: A stellar M-type spectrum with TiO molecular bands.
\item {\it NSER 10459}: Absorption-line Galaxy with $z_a=0.324$ in agreement with the previous determination by BTK. 
\item {\it NSER 10953}: Emission lines [OII]$\lambda$3727, [OIII] $\lambda$5007 and absorption features CaII K and H, H$\eta$ and H$\theta$ agree with a redshift $z=0.148$. NELG.  
\item {\it NSER 12399}: $z_e=0.425$ based on [OII]$\lambda$3727, H$\beta$ and [OIII]$\lambda$5007. $z_a=0.424$ based on CaII H, H$\eta$ and H$\theta$. NELG.  
\item {\it NSER 12472}: No emission features. Tentative absorption redshift $z_a=0.094$ based on CaII K and H.
\item {\it NSER 13310}: Quasar with $z_e=2.53$ based on broad emission-lines Ly$\alpha$, NV$\lambda$1240, CIV$\lambda$1549 and CIII] $\lambda$1909.
\item {\it NSER 13571}: We confirm the tentative redshift determination by BTK $z=0.317$, based mainly on [OII]$\lambda$3727 emission and CaII H and K absorption. NELG.
\item {\it NSER 13732}: A single broad emission-line at $\lambda_o=4675$, tentatively assigned to MgII$\lambda$2798, with a resulting redshift $z_e=0.671$.
\item {\it NSER 14260}: Galaxy without emission features. Tentative absorption redshift $z_a=0.248$ based on CaII K and H, and on CH-G band $\lambda$4304.4.
\item {\it NSER 14264}: Narrow-line AGN. $z_e=0.286$ based on [OII]$\lambda$3727, [OIII]$\lambda$4959 and $\lambda$5007. $z_a=0.287$ based on CaII K $\lambda$3934 and H $\lambda$3969. Approximate agreement with previous determination by BTK ($z=0.287$).
\item {\it NSER 15465}: Broad-line AGN with $z_e=0.528$ based on broad MgII$\lambda$2798 and forbidden [OII] $\lambda$3727.
\item {\it NSER 16338}: Very weak continuum with strong narrow emission lines, [OII]$\lambda$3727, H$\beta$, [OIII]$\lambda$4959 and $\lambda$5007. A Seyfert 2 galaxy with $z_e=0.252$. This object has been observed with both the WHT and TNG  (see figures \ref{Fig1e} and \ref{Fig2}).
\item {\it NSER 16710}: NELG. $z_e=0.439$ based on [OII]$\lambda$3727. $z_a=0.440$ based on CaII H and K.
\item {\it NSER 17475}: Broad-line AGN, with a single feature at $\lambda_o$=4352, which we identify with MgII $\lambda$2798, with a resulting redshift $z_e=0.555$.
\end{itemize}

\section{Discussion}
\begin{figure}
\centering
\resizebox{\hsize}{!}{\includegraphics{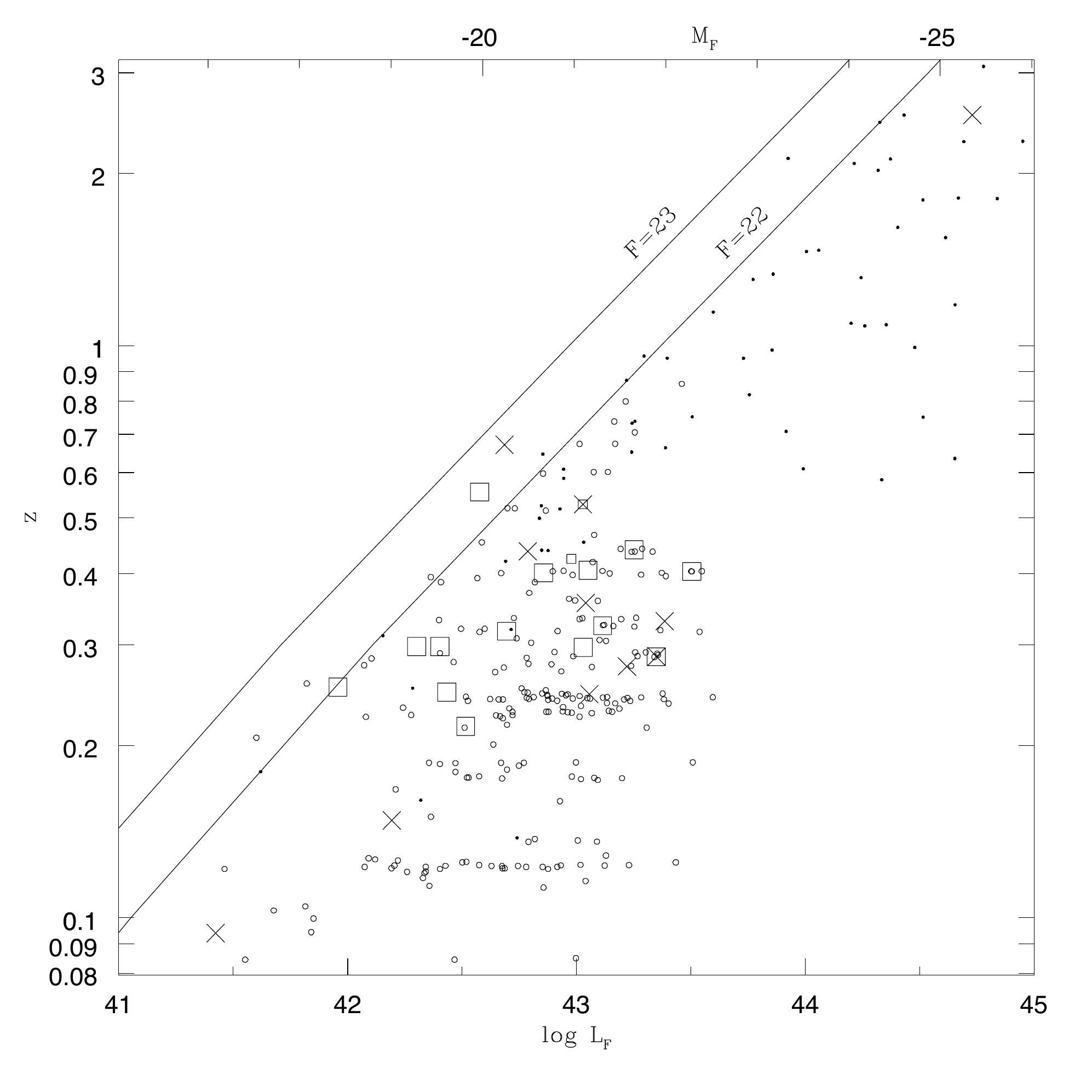}}
\caption{Objects of  SA 57 in the $L_F-z$ plane . From the present survey: large squares (extended variable candidates), small squares (point-like variable candidates), crosses (X-ray selected candidates).  Objects with known redshift from the KPNO survey \citep{munn97} are shown as:
small empty circles (extended objects), small dots (point-like objects). Lines of constant  magnitudes F=22 and F=23 are also shown.}
         \label{Fig3}
\end{figure}

%- distribuzione nel piano L-z Fig 2
% On the basis of the above data we assignn object to different classes, taking also into account the information.
%- per gli oggetti gia osservati otteniamo conferme dei redshift
Figure \ref{Fig3} shows the distribution in the luminosity-redshift  ($L-z$) plane 
of all objects  with known redshift in the field of SA 57, including 
data from the literature and the objects of the present spectroscopic campaign (larger symbols);  this contains candidates  from both the X-ray catalogue of \citet{trev07} and the lists of variable objects of \citet{t89,t94} and BTK.
The luminosity reported is computed in the $F$ band, without K-correction. Galaxy redshifts are taken from the KPNO redshift survey of SA 57 by \citet{munn97}. 
Crowding of points around $z\sim 0.125$   corresponds to a known galaxy overdensity \citep{koo84,koo87} and at  $z \sim 0.24$ corresponds
to the cluster Zw 1305.4+2941 analyzed in  \citet{kkntv88} and \citet{gas07}.
The variability-detected extended objects populate a limited region of the $L-z$ plane. This is consistent with the expectation, since
the optical survey is limited to $F \la 22$.  The galactic component becomes fainter and fainter at high $z$, and  is swamped by the nuclear luminosity for $L_F  \ga 3\times 10^{43}$ erg  s$^{-1}$, thus appearing as point-like.
It should be noted, however, that this does not reduce the interest of variability detection because: i) the main aim of the present survey is to detect faint AGNs at low redshift to verify the possible dependence 
of the evolution on intrinsic luminosity; ii) variability can detect also AGNs at higher redshift, though they appear as point-like and thus are detectable also by colour techniques. 
X-ray candidates span a wide range of redshift and luminosity ($\Delta \log L_F \sim 4$), from relatively faint X-ray sources like starburst galaxies,  to bright QSOs.
Despite the spectroscopic campaign being still incomplete, so that a discussion of the LF evolution at low luminosity is still unfeasible, we find some new interesting
objects indicating that the low luminosity part of our sample of AGN candidates consists of a mix of different object types.
In fact, while at higher luminosities ($L_F \ga 10^{43}$ erg s$^{-1}$)  most objects are broad line AGNs,
at lower luminosities  most objects are either  narrow line  AGNs  or  NELGs.

\begin{figure}
\centering
\resizebox{\hsize}{!}{\includegraphics{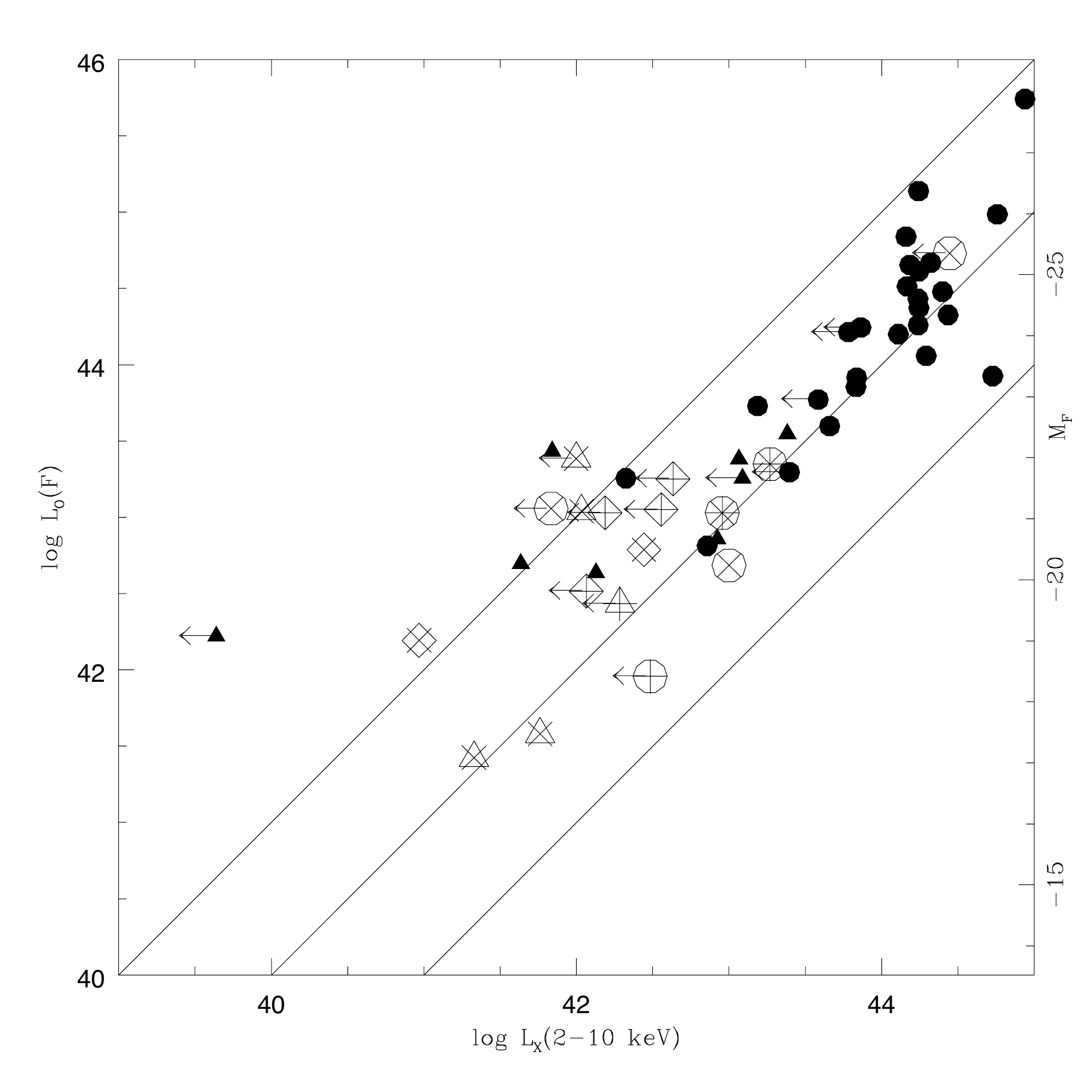}}
\caption{$L_F$ vs. $L_X(2-10 {\rm\,keV}$). 
Objects with new redshifts from the present work are plotted as open symbols with a superposed  
"$\times$" (X-ray selected) and/or a " +"
(variability selected). According to the classification in Table \ref{Tab1}, they are represented as: circles (type 1 and 2 AGNs), diamonds 
(NELGs: starbursts or LLAGNs), triangles (galactic spectra with only absorption features). 
Objects with previously known redshifts are plotted with filled symbols (as in \citet{trev07} ): circles (type 1 and 2 AGNs); triangles (galaxies with redshift from \citet{munn97}).}
         \label{Fig4}
\end{figure}

Figure \ref{Fig4} shows the optical $F$ band versus the X-ray luminosity for the objects of the present spectroscopic campaign 
plus X-ray detected objects by \citet{trev07} with previously known  redshift. The 3-$\sigma$ upper limits in the X-ray luminosity
correspond to objects not detected in the 2-10 keV band which may be either selected through optical variability and not detected in X-rays, or detected in X-rays but in a different band.

Most of the objects whose redshift has been determined in the present work  have a relatively low X-ray (2-10 keV) luminosity. Moreover, most of the objects selected through variability
are not detected in X-rays, i.e. they likely have  a low value of the X/O ratio, defined as the ratio of the  X-ray flux $f_X(2-10 \,{\rm keV})$ and the $F$-band flux $f_F$. 
They could be faint AGNs with typical nuclear X/O ($\log {X/O} \sim 1$), but with the host galaxy contributing to the observed optical luminosity.
%Notice that some objects, previously classified as NELGs \citep{munn97} 
%{\tt (triangles? ma le NELG per noi sono diamonds)} show a high X/O respect to normal galaxies 
%and their X-ray luminosity is of the order of the threshold ($L_X(2-10 {\rm\,keV}) \sim10^{42}$erg s${-1}$ usually assumed
%to separate star-burst galaxies from AGNs, or even $\sim 10$@ times this value.
Other objects, detected in X-rays, show optical spectra without emission lines, consistent with normal galaxies and
luminosities  $L_X(2-10{\rm keV}) \sim  10^{42}$  erg s$^{-1}$. Thus   they
can be classified as  X-ray Bright Optically Normal Galaxies (XBONGs) described by  \citet{fior00} and \citet{com02a,com02b}.
Different scenarios have been proposed to interpret  these objects: i) selection effects hampering line detection \citep{hor05}: ii) heavy nuclear absorption \citep{com02a}; iii) heavy extra-nuclear absorption by the  host galaxy dust \citep{rig06}; iv) strong dilution by the host galaxy light \citep{geor05}; v) radiatively inefficient accretion flow in low luminosity active nuclei \citep{yua04}. 
However all the above interpretations are based on the presence of an active nucleus.
In particular \citet{yua04} postulate the existence of a transition radius in the accretion disk, below which a radiatively inefficient accretion flow (RIAF) occurs.  According to \citet{ho99} a RIAF can also explain the absence of a big blue bump in the spectra of LINERs \citep[see however][]{mao07}.  This  suggests a relation between the two classes of objects in spite of the differences in their optical and X-ray properties.
The completion of the spectroscopic follow-up of our candidates will provide further 
data to investigate this issue.

Variability detected narrow emission line objects deserve further discussion. 
Let us consider first the objects we classified as Seyfert 2s on the basis of their high [OIII]/H$\beta$ ratio.
According to the classic unified model
\citep{anto85} the nuclear component should be hidden by the absorbing torus. 
At the same time the size of the narrow line region is such that line variability should be strongly reduced.
Thus, in our case,  the origin of variability is unclear. Notice  that variability in  some type 2 objects has been observed by \citet{kle07}. A possible explanations might  be that the broad line region is not obscured, but intrinsically lacking, so that the variable continuum can be seen \citep{gho07}. Moreover  some objects exhibit extreme spectral variations such that they appear of different type depending on the observing epoch (see for instance \citet{cze04} and refs. therein) as was already noted in the case of NSER 4326 (104326 in BTK). 
An object of this type could be  NSER 16338, which was selected on the basis of its variability during the photometric campaign in the years
1974-1989, when its magnitude was $B \sim 22.7$ and,  on the basis of the present observations (April 2006), shows a very low optical continuum with only strong emission lines.
Its spectrum is shown in both Figures \ref{Fig1e} and \ref{Fig2}, as observed with WHT and TNG respectively.
It would be interesting to monitor  this object  to detect possible long time scale spectral variations.

Part of the objects in Table \ref{Tab1} are generically classified as NELGs  since we do not observe [NII]/H${\alpha}$,  while [OIII]/H$\beta$
is not large enough to indicate the AGN character. These objects could be either starburst galaxies, or LINERs or transition objects (TOs).
However, we can look at their variability. Among the 10 NELGs in Table \ref{Tab1},  only 2 were below the variability selection threshold
and were selected solely on the basis of their X-ray emission. The other 8 NELGs were selected through their variability, and this is an evidence
in favour of their AGN character. In fact \citet{mao05}, on the basis of {\it Hubble Space Telescope} monitoring of a sample of LINERs,
conclude that UV variability is detected in most, if not all,  objects of this type. 
Moreover they conclude that this {\it murmur} of a sleeping black hole cannot be entirely explained by  the variability of luminous stars but it implies the presence of a non-stellar component. According to  \citet{cid04} and \citet{flo06}, even if the AGN is present in LINERs, often its luminosity is insufficient to explain the observed emission-line intensity. In any case our results on variability selected objects favour the presence of low luminosity AGNs in at least some narrow emission line galaxies. 

\bigskip\noindent
In summary :
\begin{itemize}
\item{We have obtained the spectra of a composite sample of variability-selected and X-ray-selected
AGN candidates;}
\item{some candidates show typical AGN spectra;}
\item{one variability selected object has strong emission lines but undetected continuum, suggesting  extreme spectral variation;}
\item{2 objects are classified as XBONGs;}
\item{most of the other objects are classified as NELGs, i.e. starbursts or LINERs;}
\item{most NELGs were variability selected suggesting that they are LINERs hosting a LLAGN.}
\end{itemize}

We are continuing our spectroscopic campaign in SA57 with various telescopes, and we plan to obtain spectra of a larger sample of faint AGN candidates selected both through variability and X-ray emission. 
This will allow us to evaluate the relevance of the different selection effects 
in the study of the evolution of the faint end of the AGN luminosity function. 
A collection of a larger sample of variability-selected LINERs will provide a better understanding of the role of the hosted AGNs.

\begin{acknowledgements}
We thank A. Cavaliere for useful comments. We acknowledge partial support of Agenzia Spaziale Italiana and Istituto Nazionale di Astrofisica by the grant ASI/INAF n. I/023/05/0. 
\end{acknowledgements}

\end{document}